\documentclass[twocolumn]{aa} % for a paper on 1 column  
\usepackage{natbib,twoopt}
\usepackage[varg]{txfonts}
\usepackage{graphicx}
\defcitealias{Ahumada_2007}{AL07}
\defcitealias{Ahumada_1995}{AL95}
\defcitealias{Vaidya_2020}{V20}
\defcitealias{Rain_2020}{R20}
\defcitealias{Cantat-Gaudin_2018}{CG18}
\defcitealias{Cantat-Gaudin_2020}{CG20}
\defcitealias{Dias_2002}{DAML02}

\begin{document}

   \title{A new, Gaia-based, catalogue of blue straggler stars \\in open clusters}

   \author{M.J. Rain
          \inst{1}
          \and
          J.A. Ahumada\inst{2}\fnmsep
          \and
          G. Carraro\inst{1}}

   \institute{Dipartimento di Fisica e Astronomia, Universita’ di Padova, Vicolo Osservatorio 3, I-35122, Padova, Italy\\
              \email{mariajoserain@gmail.com}
         \and
            Observatorio Astron\'{o}mico, Universidad Nacional de C\'{o}rdoba, Laprida 854, X5000BGR, C\'{o}rdoba, Argentina\\
          %   \email{c.ptolemy@hipparch.uheaven.space}
           %  \thanks{The university of heaven temporarily does not
            %         accept e-mails}
             }

   \date{Received September 15, 1996; accepted March 16, 1997}

% \abstract{}{}{}{}{} 
% 5 {} token are mandatory
 \abstract
 {Blue straggler stars are exotic objects present in all stellar environments whose nature and formation channels are still partially unclear. They seem to be particularly abundant in open clusters (OCs), thus offering a unique chance to tackle these problems statistically.}
 {We aim to build up a new and homogeneous catalogue of blue straggler stars (BSS) in Galactic OCs using \emph{Gaia} to provide a more solid assessment of the membership of these stars. We also aim to explore possible relationships of the straggler abundance with the parent cluster's structural and dynamical parameters. As a by-product, we also search for possible yellow straggler stars (YSS), which are believed to be stragglers in a more advanced evolution stage.}
  {We employed photometry, proper motions, and parallaxes extracted from \emph{Gaia}~DR2 for 408 Galactic star clusters and searched for stragglers within them after performing a careful membership analysis.}
  {The number of BBS emerging from our more stringent, selection criteria turns out to be significantly smaller than in previous versions of this catalogue. OCs are therefore not a preferable environment for these kinds of stars anymore.
  In addition, we found that BSS start to appear in clusters with ages larger than log(t) $\sim$ 8.7 and are therefore absent in very young star clusters.}
  {The present catalogue supersedes the previous ones in several ways: membership assessment, number of stragglers found, and so forth. The new list includes 897 BSS and 77 YSS candidates in 408 OCs.
  We expect this catalogue to be the basis for a new round of studies of BSS and YSS.} 
\keywords{star clusters and associations: general --- 
 stars: blue stragglers}

 \maketitle
%
%-------------------------------------------------------------------

\section{Introduction}

Blue straggler stars (BSS) were initially identified by \cite{Sandage_1953} in the colour-magnitude diagram (CMD) of the globular cluster M3, appearing as an extension of the cluster main sequence, blueward and above the turnoff (TO).  
Their presence poses a challenge for standard single-star evolution theory since stars with masses higher than that of the cluster TO  should have evolved into the white dwarf regime long ago. 
At present, these exotic stars have been identified   in essentially all stellar systems:  globular clusters (GCs,  \citealt{Piotto_2004,Salinas12}), dwarf galaxies (\citealt{Momany_2015}), open clusters (OCs,  \citealt{Milone_1994, Ahumada_1995, Ahumada_2007}), and  the field population of the Milky Way \citep{Preston_2000, Santucci_2015}. 
 BSS are identified according to their position in a CMD: Stars that follow the extension of the main sequence brighter than the TO are considered to be blue stragglers. In terms of magnitude, these stars typically spread from the TO to about two magnitudes brighter; in some clusters, however, the brightest BSS can be as much as 3 magnitudes above the TO. The low-luminosity limit is not always clear. For example, in OCs, stars are bluer and fainter than the TO on the zero-age main sequence (ZAMS), so they appear younger and, therefore, are  BSS -- for example, this includes stars 1366 and 8104 in NGC~188 \citep{Geller_2008} or the ones recently identified in several OCs by \cite{Leiner_2021} using \emph{Gaia}~DR2. 
Additionally, BSS cover a broader range in colour than  normal main sequence stars. Sometimes, stars located between the main sequence and the giant branch  (Hertzprung gap) are also identified as BSS, or rather evolved blue stragglers `yellow straggler stars', YSS). 
 Reviews on the topic can be found in \cite{STRYKER_1993} and
 \cite{bof15}.
By definition, BSS  are expected to be more massive than the mean stellar mass of the host cluster.  Their masses  are usually derived from  evolutionary tracks for  main sequence stars, which do not include  modifications to the standard theory of stellar evolution. Particularly in OCs, the photometric masses of some BSS are very different from those  derived directly from their binary orbits (e.g. \citealt{Sandquist_2003}).
Today it is widely accepted that a straggler started as a normal, main sequence star that has been `rejuvenated' by acquiring extra mass. The new, more massive star may reach an accordingly higher luminosity that can place it  above the CMD TO.
This increase in mass may be produced
via two, non-exclusive mechanisms:   mass-transfer in a close binary \citep{McCrea_1964} or dynamically induced stellar collisions and mergers \citep{Hills_1976, Davies_1994}. In the case of GCs, it has been suggested that each formation channel is favoured in different regions of the host cluster -- for example, collision-induced BSS mergers would be mainly active in the dense cores \citep{Verbunt_1987} -- and that both might be at work simultaneously  \citep{Ferraro_1993, Zaggia_1997, Mapelli_2006}. These scenarios, however, have not always been successful in explaining the observations. For example, BSS as mergers of previously normal main sequence stars have proposed where the progenitor of the straggler is formed in a hierarchical triple system as a result of mechanisms such as the Kozai-Lidov \citep{Ford_2000, Perets_Fabrycky_2009, Naoz_2014} or the angular momentum loss induced
by  magnetic stellar winds in a close binary. This last mechanism
would be responsible for at least one-third of the BSS in OCs older than 1~Gyr \citep{Andronov_2006}.
In  old, populated and still not very dense,  dynamically evolved  OCs, one would expect binary evolution to be the predominant mechanism
\citep{Mathieu_2015}.

Significant advances have been made concerning binarity among BSS in OCs, based on long-term spectroscopic monitoring by  \citet{Mathieu_2015}. They found a binary fraction of 70 \% in the blue straggler population, compared to about 25\% in the normal main sequence stars; this presence of binaries shows that internal dynamical processes have not induced global stellar mergers  or collisions in old OCs. Furthermore, \cite{Gosnell_2014, Gosnell_2015} detected and characterised some white dwarf companions to BSS in NGC~188, providing direct evidence for mass transfer events. We have less information regarding the binary properties of BSS in GCs. 

In stellar clusters,  BSS appear in general more centrally concentrated than the normal stellar populations. Given that they are closely related to binary evolution, it is expected that BSS should suffer mass segregation similar to the cluster's binary
population and move to the centre of the cluster potential.
In the 1990s, early studies gave the first hints for BSS showing  this effect in GCs and some OCs \citep{Aurerie_1990, Mathieu_1986}. Later studies revealed the existence of a bimodal radial distribution in these systems with a peak of stragglers at the centre, followed by a minimum at a radius $r_\mathrm{min}$, and a rise again in the outskirts of the cluster:  for the models of GCs, this was done by \cite{ Mapelli_2006} and \cite{Milone_2012}; in individual GCs, this was carried out by \cite{Lanzoni_2007} for M55 and \cite{Sabbi_2004} for NGC~6752; and for individual OCs,  this was performed by \cite{Geller_2008} for NGC~188, \cite{Carraro_2014-1} for Melotte~66, \cite{Bhattacharya_2019} for Berkeley~17,  and \cite{Rain_2020b} for Collinder~261. The importance of this $r_\mathrm{min}$ lies on its correlation with the dynamical age of the cluster (GCs: \citealt{Ferraro_2012}; OCs: \citealt{Vaidya_2020}).

The study of BSS  in OCs is a current high-interest subject since they give information about the number of binary systems and binaries' roles in cluster evolution.
It  enables an in-depth analysis of blue straggler characteristics such as frequency, orbital parameters, and masses, which are the most important diagnostic tools for determining their origin. Moreover, because of their  sparse nature,  OCs are good laboratories to study the nature and formation of BSS, especially with spectroscopy. So far, the most extensive survey of BSS candidates in OCs has been performed by \citet{Ahumada_2007}, hereafter AL07, which took into account the
 bibliography up to 2005. This catalogue was, in turn,  a revision of the previous work by \citet{Ahumada_1995}, hereafter AL95, and it lists a total of 1887 blue straggler candidates in 427 OCs of all ages. \citetalias{Ahumada_2007} found that: i) stragglers are present in clusters of all ages; ii) the BSS show a remarkable degree of central concentration,  and iii) the fraction of BSS increases with the richness and age of the cluster. 
Drawbacks of AL07 are  the lack of homogeneity of the open cluster data available at the time it 
was published, and that the straggler candidates are mostly of an uncertain membership. 
Thus, while useful, these compilations are  not reliable enough to allow the derivation of statistical properties of BSS.
Today, an improvement in the selection of BSS  has become possible thanks to the second data release of the \emph{Gaia} mission (\citealt{Gaiacoll2018}, hereafter \emph{Gaia}~DR2),  which permits better discrimination of genuine BSS from field stars by using high-quality proper motion and parallax information. Using {\it Gaia}~DR2, we can determine accurate membership and, consequently, raise the study of BSS in OCs to  much more solid statistical grounds for the  first time.

This paper presents a new catalogue of blue straggler stars in a large sample of OCs.  
It is  based on the inspection of the CMDs of 408 clusters with a membership characterisation provided by the \emph{Gaia}~DR2 astrometric solution. The paper is organised as follows.
In \S~\ref{sec:search_bss}, we describe the data and the methodology used to identify the BSS and YSS. 
In \S~\ref{sec:catalogueo} we give a detailed explanation of the contents of the catalogue itself, while in \S~\ref{sec:estadistica} we provide some general statistics. In \S~\ref{sec:conclusions} we summarise and give the conclusions of this work.
%----------------------------------------------------------------------

\section{Search for blue stragglers in open clusters}\label{sec:search_bss}
%----------------------------------------------------------------------

This work is based on identifying blue straggler candidates in  colour–magnitude diagrams of galactic OCs of all ages. 
This is a new compilation,  for which we  made use of the recent \emph{Gaia}~DR2 survey. For this catalogue, we also searched for YSS and introduced some methodological improvements.
This section is devoted to explaining how the compilation was performed, while the catalogue's proper description is in \S~\ref{sec:catalogueo}.

%----------------------------------------------------------------------
\subsection{List of clusters}\label{subsec:list_clusters}
%----------------------------------------------------------------------

From the original list of \citetalias{Ahumada_2007}, we extracted 389 clusters, to which we added 19 recently discovered ones with the \emph{Gaia} mission, for a grand total of 408; they  all appear in \cite{Cantat-Gaudin_2018}.
This way,  we  dealt with a sample of clusters very similarly to those in the previous catalogues. We note, however, that the list of clusters can be easily expanded. 
Table~\ref{tab:clusters_notinR20} lists all  clusters that are included in AL07 but not here; they are not considered to be bona~fide clusters, mainly according to a non-compliance with two simple conditions using the \emph{Gaia}~DR2 astrometric solution: i)  their proper-motion dispersions correspond to a physical velocity dispersion inferior to 5~km~s$^{-1}$; and  ii) their observed proper-motion dispersions are smaller than  1~mas~yr$^{-1}$ (three times the \emph{Gaia} measurement errors; also see \S~\ref{subsec:gaialimitaciones}).  For  more details, the reader is referred to \cite{Cantat-Gaudin_2020}.
Furthermore, our sample does not include two  well known and close clusters, Hyades (Melotte~25) and Coma (Melotte~111), given their large extension on the sky and the more sophisticated membership determination technique required to identify their BSS properly. 
 On the other hand,
 the clusters listed for the first time  are in Table~\ref{tab:clusters_new}.

\addtolength{\tabcolsep}{11pt}    
\begin{table}[!ht]
{\small
\centering
\caption{Open clusters with entries in AL07, but not included in this work.}
\label{tab:clusters_notinR20}
  \begin{tabular}{l l l} 
  \hline
  Berkeley 42 & Berkeley 64 & Berkeley 66 \\
    Bochum 1    & Bochum~2     &  Bochum 7 \\
    Bochum 10 & Bochum 14 &Collinder~96 \\
    Collinder~97 & Collinder~121 & Collinder~223\\
    Collinder~228    &  Coma     & Dolitze 25\\
     Feinstein~1   &  Hogg 16    & Hogg 22  \\
     Hyades & IC 2944 & NGC 133 \\
    NGC 1252 & NGC~1931      & NGC~1976\\
    NGC 2175 & NGC 2384 & NGC 2467 \\
    NGC 3247 & NGC 6200 & NGC 6514 \\
    NGC 6530 & NGC 6604 & Pismis 24 \\
    Ruprecht 46 & Ruprecht 55 & Ruprecht 120 \\
    Stock 13 & Trumpler 24 & Trumpler 27 \\
    \hline
  \end{tabular}
  }
\end{table}

\addtolength{\tabcolsep}{4pt}
\begin{table}[!ht]
{\small
\centering
\caption{Open clusters not included in previous versions of the catalogue.}
\label{tab:clusters_new}
  \begin{tabular}{l l l} 
  \hline
  Gulliver~1  & Gulliver~4   & Gulliver~40 \\
  Gulliver~12 & Gulliver~51  & Gulliver~52\\
  Gulliver~21 & Gulliver~55  & Coin~2 \\
  Gulliver~23 & Coin~15      & Coin~35 \\
  Gulliver~27 & Trumpler~11  & Trumpler~20  \\
  Gulliver~36 & Waterloo~7   & Pozzo~1 \\
  Gulliver~39 & & \\
  \hline
  \end{tabular}
}
\end{table}

%-----------------------------------------------------------------------
\subsection{Milky Way open clusters with \emph{Gaia}~DR2 and membership criteria}
\label{sec:miembros}
%-----------------------------------------------------------------------

The survey \emph{Gaia}~DR2 provides a  precise astrometric solution (RA, DEC,  $\mu_{\alpha^\ast}  (=\mu_{\alpha}\cos \delta)$, $\mu_{\delta}$, and $\varpi$) for $\sim 1.7$ billion objects. In addition,
 it contains photometric data in the $G$-band for all sources, while data in the $G_\mathrm{BP}$- and $G_\mathrm{RP}$-bands are available for 80\% of the sources ($\sim1.4$ billion objects). The limit for faint magnitudes is  $G\approx21$, while the bright limit for $G$ is around 13. The calibration uncertainties reached on the individual observations are 2, 5, and 3~mmag for $G$, $G_\mathrm{BP}$, and $G_\mathrm{RP}$, respectively. 
The sources present a median uncertainty in parallax and position of about 0.04~mas for bright sources ($G< 14$ mag), 0.1~mas at $G = 17$~mag, and 0.7~mas at $G = 20$ mag. In the proper motion components, the corresponding uncertainties are 0.05, 0.2, and 1.2~mas~yr$^{-1}$. The astrometric solution, the photometric contents and validation, and the properties and validation of radial velocities are described in \cite{Lindegren_2018}, \cite{Evans_2018}, and \cite{Katz_2019}. The homogeneity and quality of \emph{Gaia} data allow us to reach an unprecedented level of detail in CMDs, particularly in the context of OCs, where accurate parallax information has   often been lacking. 
 
We took advantage of the selection of cluster members carried out by \cite{Cantat-Gaudin_2018}, hereafter CG18, and updated by \cite{Cantat-Gaudin_2020}, hereafter CG20. The last version contains 1481 clusters and provides coordinates, proper motions,  parallaxes, and distances for all of them.
To assign membership probabilities ($P_\mathrm{memb}$),  \citetalias{Cantat-Gaudin_2018} and \citetalias{Cantat-Gaudin_2020} applied the membership assignment code UPMASK\footnote{Unsupervised  Photometric  Membership  Assignment in Stellar Clusters} to the \emph{Gaia}~DR2 data in  the field of each cluster. This field has  a radius twice as large as the value  $r_\mathrm{DAML02}$ reported by \cite{Dias_2002}, with proper motions within 2~mas~yr$^{-1}$, and  a parallax within 0.3~mas, of those of the cluster centroid ($\mu_{\alpha^*}$, $\mu_{\delta}$, $\varpi$). Since the uncertainties  of $G$ at $\sim21$~mag reach 5 mas~yr$^{-1}$ for proper motions, and 2 mas for parallaxes,  the Cantat-Gaudin catalogue includes only stars with $G \leq 18$, which corresponds to  typical uncertainties of 0.3 mas~yr$^{-1}$ and 0.15 mas in proper motions and parallax. See our discussion in 
\S~\ref{subsec:gaialimitaciones}.

%-----------------------------------------------------------------------
\subsection{Identification of the blue stragglers}
\label{sec:identificacion}
%-----------------------------------------------------------------------

\begin{figure}
\centering 
\includegraphics[width=\linewidth]{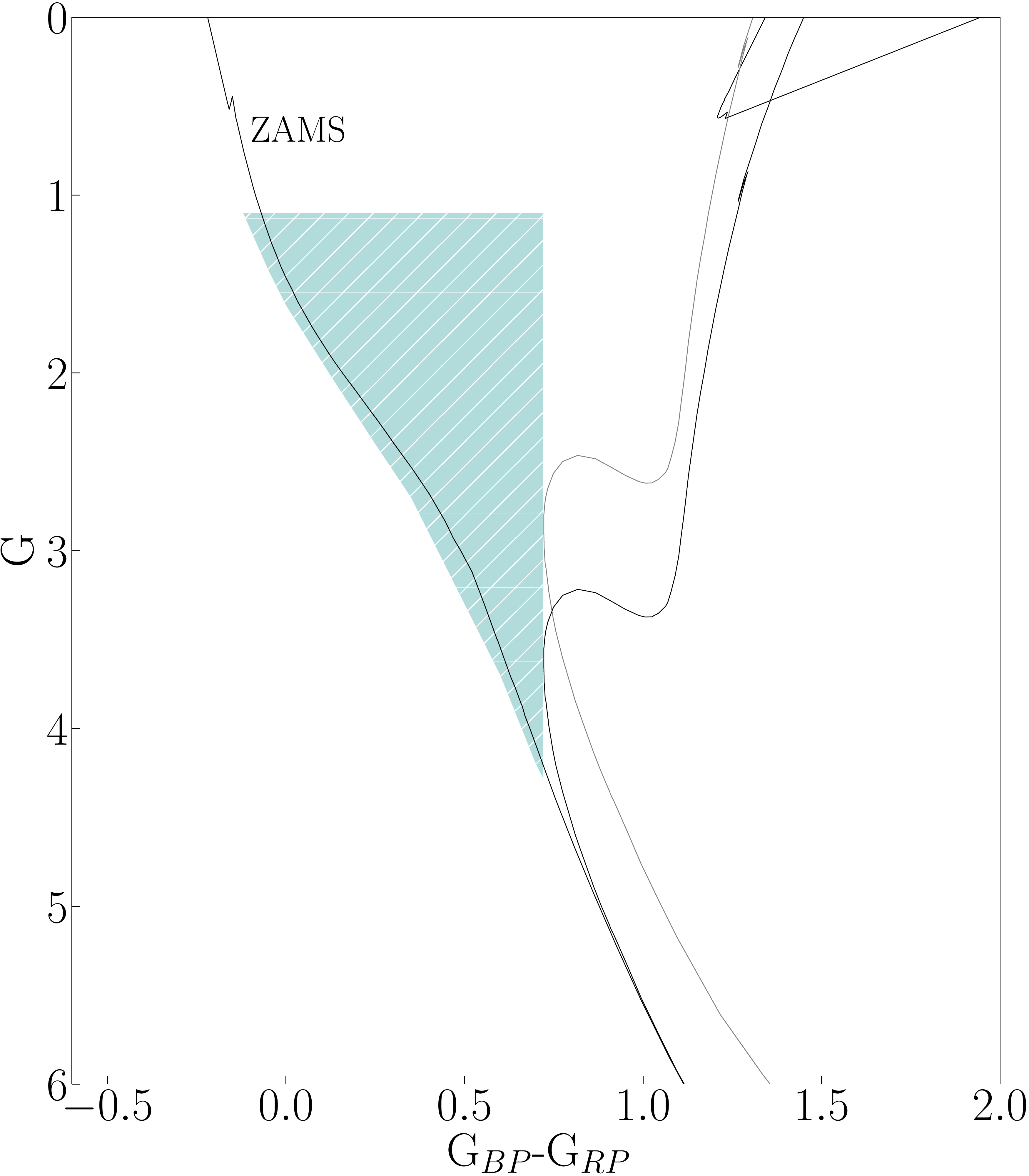}
\caption{Schematic colour–magnitude diagram for an old open cluster;
the isochrone corresponds to $\log(\mathrm{age}) = 9.8$. The blue straggler area is
shaded in blue. The grey line indicates the equal-mass binary loci.}
    \label{fig:cmd_example_bs}
\end{figure}

Formally, a blue straggler was defined from its position in a CMD, that is, it is bluer and brighter than the TO, on or near the parent cluster ZAMS.  Following \citetalias{Ahumada_1995}, \citetalias{Ahumada_2007}, and \citetalias{Rain_2020}, we  assumed a cluster star to be a blue straggler candidate if it appears on a specific area of the CMD (see Figure~\ref{fig:cmd_example_bs}). This region is bounded on the blue side by the ZAMS and the red side by the TO colour and the binary sequence. The lower limit corresponds to the magnitude at which the observed sequence of the cluster detaches from the ZAMS. In principle, we did not adopt a bright limit (but see \S~\ref{subsec:masivos}).

\begin{figure}
\centering 
\includegraphics[width=\linewidth]{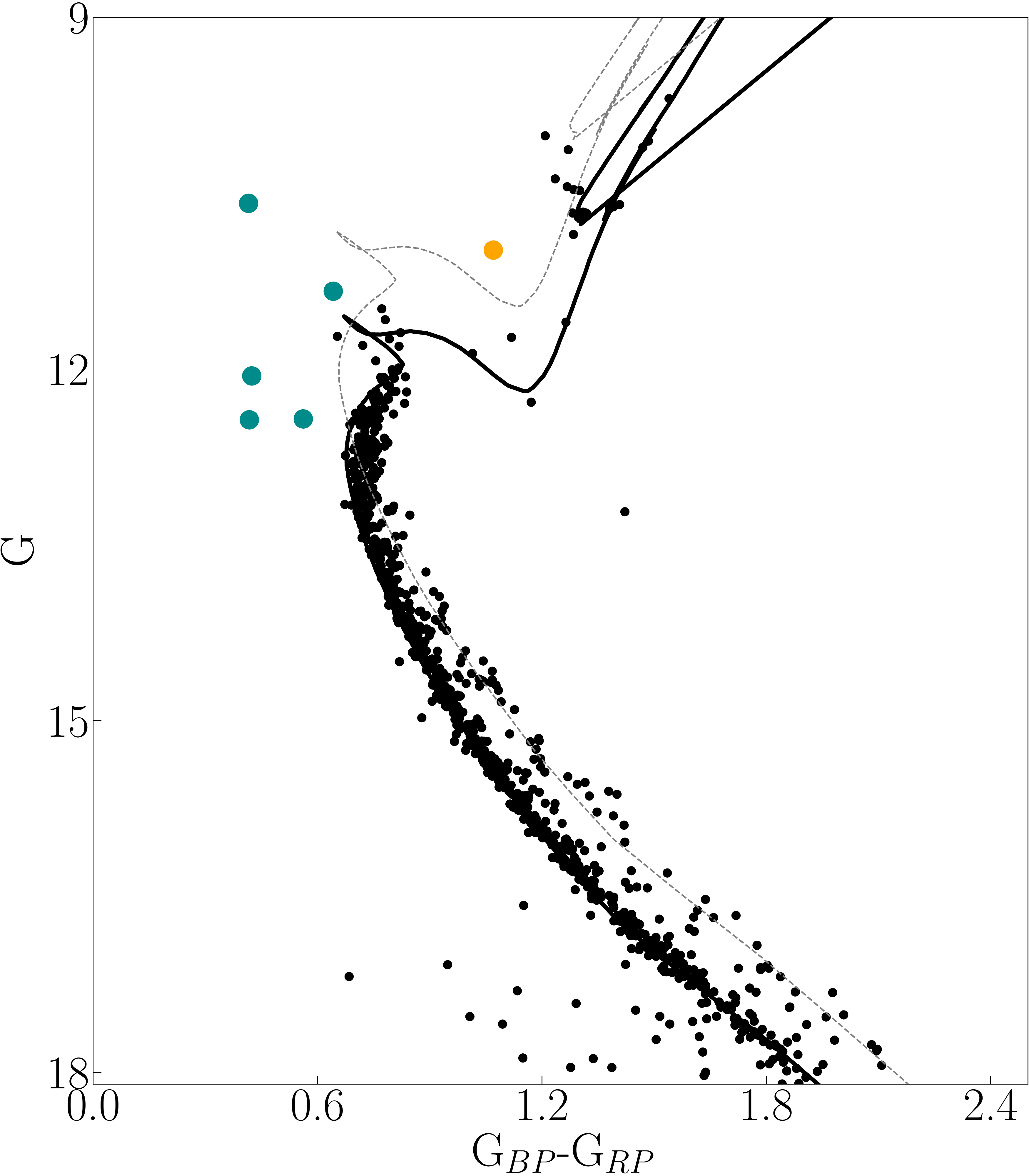}
\caption{CMD of the open cluster IC~4651, built from \emph{Gaia}~DR2 photometry.  Black filled circles are all the stars from \citetalias{Cantat-Gaudin_2018} with $P_\mathrm{memb} \geq 50$\%.  Blue filled circles and yellow filled circles are the catalogued blue and YSS, respectively. The black solid line represents the \textsc{PARSEC} isochrone  of solar metallicity and $\log(\mathrm{age}) = 9.27$ \citep{Bressan_2012},  set at $E (B - V) = 0.36$ and $(m-M)_0 = 9.82$. The second curve is the same isochrone displaced 0.75~mag, which helps constrain the region of the binaries. See also Table~\ref{tab:example_BSS}.
    \label{fig:cmd_real}}
\end{figure}

For more details, we refer the reader  to the analogous definitions in \citetalias{Ahumada_1995} and \citetalias{Ahumada_2007}, which are the models for the present one.
To identify the BSS candidates, the procedure for each cluster was the following; we note that the reader may also find Section~2.2.1 of \citetalias{Ahumada_2007} useful:
\begin{enumerate}

    \item The photometric data for all  \citetalias{Cantat-Gaudin_2018} members (\S~\ref{sec:miembros}) with probabilities of $P_\mathrm{memb}\geq 50$\%  were plotted in a $G$ versus\ ($G_\mathrm{BP}-G_\mathrm{RP}$)  diagram.
    \item An approximate matching of a \textsc{PARSEC} theoretical isochrone  \citep{Bressan_2012}, with the \emph{Gaia}~DR2 passbands of \cite{Evans_2018},  was then performed on the main sequence and TO, and eventually on the red giant branch (RGB) and red clump if present. Cluster parameters, such as metallicity,  extinction, and the age of DAML02, were first chosen. When the cluster was not listed in DAML02 or when the matching of the isochrone was unsatisfactory, parameters from \cite{Bossini_2019} and \cite{Monteiro_2019} were used instead. However, the distance  was always taken from CG18.  Sometimes a refinement of the parameters was necessary; in such cases, we kept the distance and extinction values fixed and varied the age -- that is to say, we changed the isochrone, adopting the solar value when no metallicity  was listed in the literature.  Based on these parameters' variation, we picked the appropriate PARSEC isochrone by eye, from a grid spaced 0.05 in $\log(\mathrm{age})$. These changes are reported in the catalogue Notes (see \S~\ref{subsec:notes}).
    \item The equal-mass binary loci -- obtained by displacing the isochrone by 0.75~mag upward  -- and the ZAMS  were also plotted. The binary loci help constrain the region expected to be populated by binaries made of normal main sequence TO stars.
     \item Finally, the stragglers' candidates were singled out in the corresponding region of the CMD. As an example, Figure~\ref{fig:cmd_real} shows  this selection in the CMD of IC~4651. 

\end{enumerate}

The aforementioned procedure was executed, taking the following considerations  into account. First, stars located slightly towards the blue of the ZAMS were also regarded as straggler candidates. 
Second, not all the stars fainter than the cluster TO and located between the ZAMS and the isochrone were included: Only objects significantly detached from the cluster isochrone, for instance $\sim0.03$~mag as a minimum and down to  0.5--1.5~mag  at most below the TO, were listed. This magnitude cut is consistent with our limit of $G\approx 18$ for the oldest and distant clusters in our sample (see \S~\ref{subsec:gaialimitaciones} and \citetalias{Cantat-Gaudin_2018}). 
Third, although we took the TO colour for the straggler region's red limit, an arbitrary cut-off in colour was imposed for some clusters similarly as in AL07. Redder stars were considered as possible YSS (YSS, \S~\ref{sec:yellows}). 
Finally, in this work, as in AL07, stars on the `blue hook' were not regarded as stragglers, given the dependence of this feature's extension and shape with the stellar models adopted for the isochrone. In this sense, the adoption of the equal-mass binary sequence  (see, e.g. Figure~\ref{fig:cmd_real}) was beneficial. This sequence wraps and helps to  constrain the blue hook region of the cluster, when possible, since here normal photometric binaries may fall, increasing the straggler population spuriously.

\subsection{Massive stragglers}
\label{subsec:masivos}

Although we decided not to define a bright limit in the BS area (\S~\ref{sec:identificacion} and Fig.~\ref{fig:cmd_example_bs}),
there is actually a sort of frontier given by the theory of mass transfer in close binaries of 2.5~mag above the TO
\citep{Chen_2009, McCrea_1964}: This is the luminosity that would reach the complete merger of two identical (same mass) main sequence stars at the TO.
However, there is evidence of brighter, bona~fide BSS similar to, for instance, star~677 in NGC~7789   \citep{bre80}. Stars similar to this could be the result, for example, of a three-body 
interaction and subsequent merger. We flagged these stars as `massive' to identify them as
subjects of further, specific research. To locate them, we used isochrones
whose TOs are 2.5~mag brighter than those of the isochrones used to match the CMDs.
In practice, this implied adopting an upper limit of about 1-1.5~mag above the 2.5~mag frontier.

%-----------------------------------------------------------------------
\subsection{Yellow stragglers}
\label{sec:yellows}
%-----------------------------------------------------------------------

\begin{figure}[ht]
\centering 
\includegraphics[width=\linewidth]{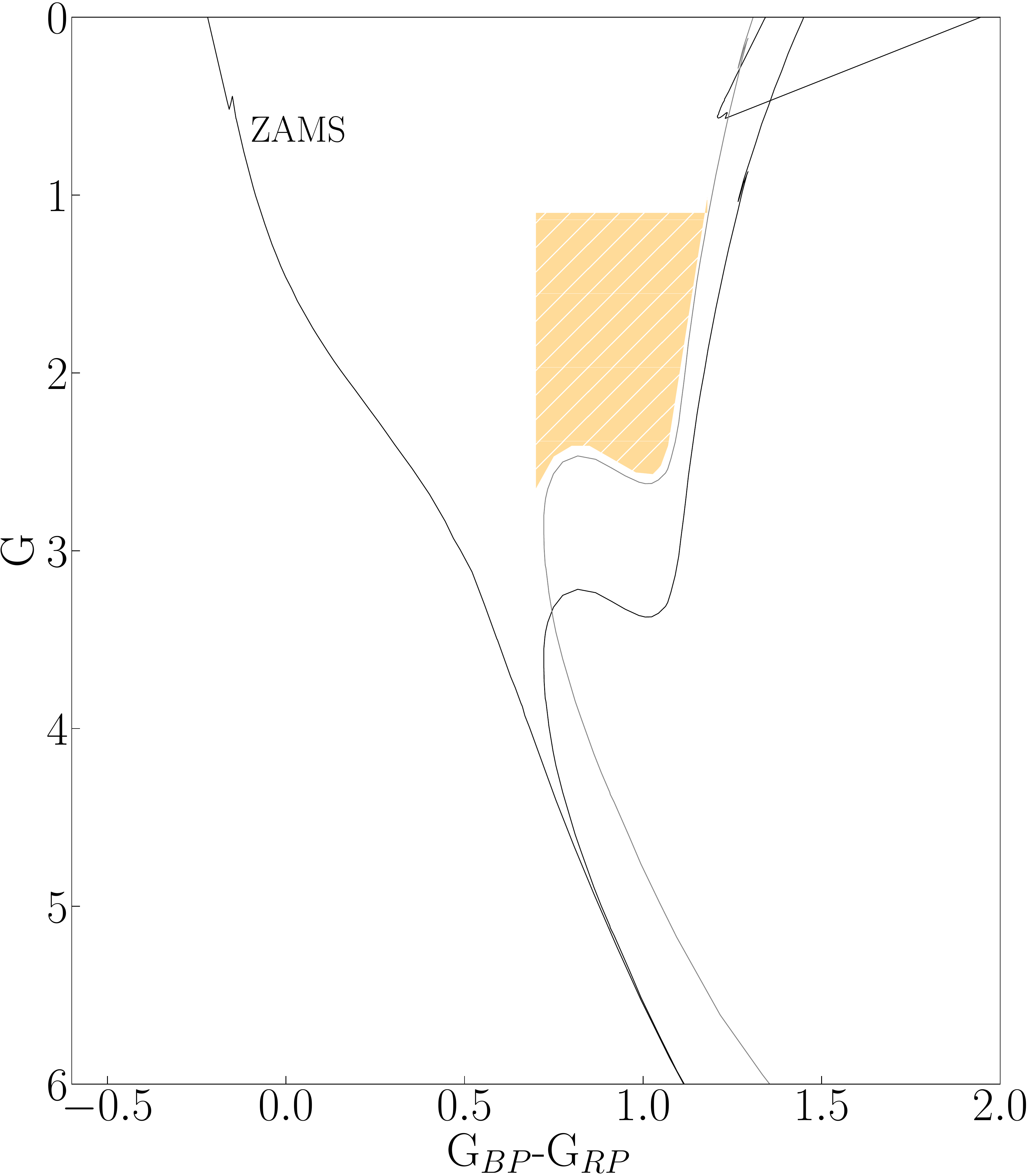}
\caption{Yellow straggler area in a schematic colour–magnitude diagram of an old open cluster ($\log(\mathrm{age}) = 9.8$). The equal-mass binary loci is also plotted.}
    \label{fig:cmd_example_ys}
\end{figure}

We name YSS the objects with colours between those of the TO and the RGB, and they are brighter than the subgiant branch  \citep{Clark_2004}.\footnote{In the literature, there is a potential source of confusion regarding the naming  of these stars since the `red straggler' term has also been systematically used for  stars lying  between the blue straggler area and the RGB (e.g. \citealt{Eggen_1983, Eggen_1988, Landsman_1997, Landsman_1998, Albrow_2001}). In this work, we chose to designate these objects as `yellow stragglers', and we encourage the community to adopt this nomenclature.} They  have been photometrically and spectroscopically identified in both open and globular clusters \citep{Silva_2014,daSilveira_2018, Rain_2020b, Martinez_2020}, and they are usually explained as  evolved BSS  -- post-main sequence stars more massive than the TO -- on their way to the RGB (\citealt{Mathieu_1990} and references therein). For example, 
 four YSS have been identified in M67,  one of them with a Helium white dwarf (WD) companion, indicating that it is an evolved straggler that formed from mass transfer,  having an RGB star as a donor \citep{Landsman_1997, Landsman_1998}. \cite{Leiner_2016} reported the first asteroseismic mass and radius measurements of the yellow straggler S1237 in M67; they argue that it might be the result of a stellar collision or a binary merger. 
The methodology followed for finding  YSS was similar to the one used for BSS (\S~\ref{sec:identificacion}).  
Figure~\ref{fig:cmd_example_ys} shows the region where these stars were searched for. This region is limited to the left by the TO colour or red limit depending on the cluster. Below and to the right by the grey equal-mass photometric binary sequence,  which corresponds to the matched isochrone that moved  0.75 to brighter magnitudes and that represents the maximum brightness expected 
for an equal mass binary at the cluster TO.  In OCs, however, the YSS regime can also contain the binaries' product of mass transfer or mergers (models of \citealt{Tian_2006} and \citealt{Chen_2008}), and they can even be `contaminated' by  objects that formed by two-member stars (e.g. a red giant-main sequence star), whose combined lights photometrically place the system within the Hertzsprung gap  \citep{Mermilliod_2007}. 

Finally, although no upper limit was defined, and considering that the CMDs can be affected by differential extinction -- which may distort the clump structure---those stars with magnitudes fainter than  the  clump and clearly differentiated from the cluster sequences were also labelled as possible YSS.

%-----------------------------------------------------------------------
\subsection{Limitations of \emph{Gaia}~DR2 photometry}
\label{subsec:gaialimitaciones}
%-----------------------------------------------------------------------

\emph{Gaia}~DR2 suffers from natural calibration problems and systematic errors, 
both for very bright and faint sources.  Here, we briefly describe how these limitations can affect the completeness of our catalogue.

For faint sources ($G\geq$  17--18~mag), problems in the background calibration and contamination from nearby sources have been reported (\citealt{Evans_2018}; \citealt{Arenou_2018}). For this reason, our catalogue contains only sources brighter than $G\approx18$, which is the limit adopted by \cite{Cantat-Gaudin_2018}. This cut in terms of distance corresponds to the de-reddened magnitude of a 3-Gyr star cluster's TO at 10~kpc.  Hence, we did not include a few  distant and old  OCs with TOs fainter than $G\approx18$   in the
catalogue because  of the virtual impossibility of matching an isochrone to a CMD without a clear TO. An example is
Berkeley~66, whose TO is at $G\sim19$.

In \emph{Gaia}, the sources described as `bronze' are photometrically non-calibrated objects, without colour measurements, and for which only the $G$ magnitude is provided. They tend to prevail in high-density stellar environments. These sources cannot be represented in a CMD, and therefore they will not be catalogued. An example is star 266 of NGC~2354, which is a $\beta$~Lyrae eclipsing binary
(V*~QU~CMa, \citealt{Lapasset_1996}).

On the other hand, the bright end  ($G\leq 13$)  is limited by photometric calibration errors. In clusters with a TO brighter than $G \approx 12$, where saturation and flux loss becomes important, and the instrument calibration is still very provisional and  problematic, some BSS may be missing. For example,  IC~2602 ($G_\mathrm{TO} \sim 4$) harbours the bright straggler  $\theta$~Carinae (HD~93030, \citealt{Walborn_1979, Naze_2008}), which does not appear in \emph{Gaia} and therefore has not been included in our list.

 \subsection{Field contamination in the straggler region}

 Usually, field stars located between us and a cluster contaminate its CMDs. This is particularly true for OCs since most of them are located low onto the disc, and many are projected towards the Galaxy bulge. 
 In fact, \cite{Carraro_2008} showed that stars belonging to the young stellar population of the Milky Way disc tend to occupy the same region on the CMD as the BSS, artificially enhancing the clusters' BSS population. Cluster members, selected as in  \citetalias{Cantat-Gaudin_2018} on an astrometric basis only, are expected to include a fraction of interlopers. 
 Disc star contamination is then one of the main reasons we decided to retain only potential members with $P_\mathrm{memb} \geq 50\%$. This criterion should enhance the probability of picking up the most likely members  \citep{Carrera_2019, Banks_2019, Yontan_2019} and providing more robust BSS statistics.
 
%-----------------------------------------------------------------------
\subsection{Differential reddening}
\label{subsec:differential_reddening}
%-----------------------------------------------------------------------

It is well known that the patchy dust distribution in the field of view towards star clusters causes  differential extinction (see, e.g. \citealt{Platais_2008}); this manifests as a broadening of the stellar sequences in CMDs. For old OCs (age $\geq 1$~Gyr), the effects of differential reddening are most noticeable in the TO and RGB morphologies. In particular, the \emph{Gaia} photometric bands are broad enough to introduce large colour differences  caused by extinction as a function of the  stellar SED. These spreads in colour can introduce some dispersion in the CMD positions, affecting the selection, especially  near the  TO. As shown by \cite{Leiner_2021}, for clusters with low reddening values ($E(G_\mathrm{BP}-G_\mathrm{RP}) < 0.3$), it is sufficient to  adopt  reddenings  from the literature and convert them to the \emph{Gaia} passbands. On the other hand, for clusters with high reddenings, individual reddening corrections are recommended.  Although a small number of clusters in our sample show high differential extinction across their field, we did not attempt to correct the photometry of the individual sources  from interstellar extinction; in these cases, a warning was included in the Notes (\S~\ref{subsec:notes}). In \cite{Rain_2020b}, we explored this effect in two OCs (Trumpler~5 and Trumpler~20). We conclude that their straggler populations did not change considerably despite their low-latitude Galactic position  and  high extinction values. We refer  the reader to that study for more information regarding the methodology used to quantify this effect in OCs using \emph{Gaia}~DR2 data.

%-----------------------------------------------------------------------
\section{The catalogue}
\label{sec:catalogueo}
%-----------------------------------------------------------------------

This new catalogue of BSS in OCs  contains the following files:  one first table with open cluster data,  a second table  listing  blue straggler data,  a third table  with  the yellow straggler compilation, and, finally, a file with notes and comments.

%-----------------------------------------------------------------------
\subsection{First table: Open cluster data}
\label{subsec:first_table}
%-----------------------------------------------------------------------

The cluster parameters were mainly taken from \citetalias{Dias_2002}, \citetalias{Cantat-Gaudin_2018}, and
\cite{Bossini_2019}. For clusters recently discovered, we used the data provided by \cite{Monteiro_2019}. The parameters listed for each cluster are as follows: the equatorial coordinates  (J2000.0), the  logarithm of the age, the reddening, the mean proper motions, the parallaxes, the distance, the number of BSS and YSS candidates with $P_\mathrm{memb} \geq 50\%$, and eventual notes (cf.\ \S~\ref{subsec:notes}). In total, the table comprises twelve columns. 
Table~\ref{tab:example_OCs} is an extract of the full Table, which  is only available
in electronic form.

\addtolength{\tabcolsep}{-15pt}
\begin{table*}[]
    \centering
        \caption{Excerpt of Table 1 of the catalogue compiling general information of the open clusters. The full version of Table 1 is only available with the electronic version of the article}
    \label{tab:example_OCs}
    \begin{tabular}{l c c c c c c l l l c c}
    \hline \hline
       Cluster   & RA  & DEC& $\log(\mathrm{age})$ & $E(B-V)$ & $\langle \mu_{\alpha^\ast} \rangle$   &   $ \langle \mu_{\delta}\rangle$ & $\langle \varpi\rangle$ &  Distance  & $N_\mathrm{BSS}$ & $N_\mathrm{YSS}$ & Notes\\
       & [degrees] & [degrees] & & & [mas~yr$^{-1}$] & [mas~yr$^{-1}$] & [mas] & [pc] & & & \\
       \hline
  NGC\_7788  & 359.179 & 61.395 & 7.26 & 0.52 & $-3.112$ & $-1.762$ & 0.300 & 3039.2 & 0 & 0 & *\\
  King\_12   & 358.265 & 61.953 & 7.85 & 0.51 & $-3.416$ & $-1.420$ & 0.296 & 3080.6 & 0 & 0 & \\
  King\_11   & 356.912 & 68.636 & 9.04 & 1.27 & $-3.358$ & $-0.643$ & 0.262 & 3433.2 & 18 & 0 & *\\
  King\_20   & 353.305 & 58.469 & 8.30 & 0.65 & $-2.686$ & $-2.585$ & 0.496 & 1903.8 & 0 & 0 &\\
  NGC\_7654  & 351.195 & 61.590 & 8.20 & 0.57 & $-1.938$ & $-1.131$ & 0.596 & 1600.1 & 0 & 0 &\\
  Berkeley\_99 & 350.260 & 71.778 & 9.50 & 0.30 & $-3.139$ & $-0.359$ & 0.137 & 6029.4 & 5 & 1 & *\\
  Mrk\_50    & 348.806 & 60.448 & 7.09 & 0.81 & $-3.465$ & $-2.560$ & 0.323 & 2837.6 & 0 & 0 &\\
  NGC\_7510  & 347.767 & 60.579 & 7.35 & 0.90 & $-3.664$ & $-2.193$ & 0.286 & 3177.5 & 0 & 0 &\\
  King\_19   & 347.053 & 60.523 & 8.55 & 0.54 & $-4.838$ & $-2.651$ & 0.343 & 2687.7 & 1 & 0 &\\
  King\_10   & 343.748 & 59.170 & 7.44 & 1.13 & $-2.722$ & $-2.088$ & 0.259 & 3478.1 & 0 & 0 &\\
  NGC\_7419  & 343.579 & 60.814 & 7.15 & 2.02 & $-2.759$ & $-1.601$ & 0.280 & 3235.9 & 0 & 0&\\
  NGC\_7380  & 341.817 & 58.125 & 7.07 & 0.60 & $-2.517$ & $-2.144$ & 0.333 & 2765.6 & 0 & 0&\\
  Berkeley\_96 & 337.478 & 55.408 & 7.60 & 0.17 & $-3.503$ & $-2.994$ & 0.252 & 3560.6 & 0 & 0 &\\
  NGC\_7261  & 335.056 & 58.128 & 8.20 & 0.88 & $-3.932$ & $-2.909$ & 0.278 & 3261.9 & 0 & 0 &\\
  NGC\_7243  & 333.788 & 49.830 & 8.00 & 0.18 & $+0.433$ & $-2.857$ & 1.116 & 873.3 & 0 & 0 & \\
  NGC\_7235  & 333.083 & 57.271 & 7.00 & 0.81 & $-2.381$ & $+2.935$ & 0.152 & 5536.9 & 0 & 0 & *\\
  NGC\_7209  & 331.224 & 46.508 & 8.53 & 0.17 & $+2.255$ & $+0.283$ & 0.820 & 1177.7 & 0 & 0 & *\\
  NGC\_7160  & 328.448 & 62.589 & 7.27 & 0.37 & $-3.472$ & $-1.378$ & 1.050 & 926.7 & 0 & 0\\
  IC\_5146   & 328.372 & 47.246 & 6.00 & 0.59 & $-2.910$ & $-2.490$ & 1.213 & 805.0 & 0 & 0 & \\
  NGC\_7142  & 326.290 & 65.782 & 9.55 & 0.35 & $-2.747$ & $-1.288$ & 0.392 & 2376.4 & 10 & 1 &*\\
  \vdots &\vdots & \vdots&\vdots &\vdots & \vdots & \vdots & \vdots& \vdots&\vdots & \vdots & \vdots  \\
%   \dots &\dots & \dots&\dots &\dots & \dots & \dots & \dots& \dots&\dots & \dots \\
       \hline 
    \end{tabular}
\tablefoot{In the table, the clusters are sorted by right ascension.} 

\end{table*}

%-----------------------------------------------------------------------
\subsection{Table 2: Blue stragglers in open clusters}
\label{subsec:second_table}
%-----------------------------------------------------------------------

In this table, we list information on every identified  straggler candidate.   In total, the table contains eleven columns. The first column gives the cluster's common name, while the second column indicates the  \emph{Gaia}~DR2 identification. Columns three to seven list the coordinates, the individual proper motions, and the parallaxes. Columns eight and nine provide the $G$ magnitude and the $(G_\mathrm{BP}-G_\mathrm{RP})$ index. Column ten indicates the membership probability by \citetalias{Cantat-Gaudin_2018}. Column eleven lists the distance from the cluster centre.  
As an example, Table~\ref{tab:example_BSS} shows the entries for the cluster  IC~4651; the full Table is only available in electronic form.

\addtolength{\tabcolsep}{-3pt}
\begin{table*}[]
    \centering
        \caption{Blue straggler candidates in the open cluster IC~4651 (see also Figure~\ref{fig:cmd_real}). 
        This is an extract of the full Table 2
        of the catalogue, which is only available in electronic form.}
    \label{tab:example_BSS}
    \begin{tabular}{l c c c c c c c c c c c c c}
    \hline \hline
       Cluster  &\emph{Gaia}~DR2 Source Id. &  RA & DEC   &  $\mu_{\alpha^\ast}$   &   $\mu_{\delta}$  &  $\varpi$ &  $G$  & $G_\mathrm{BP} - G_\mathrm{RP}$  & $P_\mathrm{memb}$ &  $r$ \\
       & & [degrees] & [degrees] &  [mas~yr$^{-1}$] & [mas~yr$^{-1}$] & [mas] & & &  &  [arcmin]  \\
       \hline
%  IC\_4651 & 5949553522087135616 & 261.231 & $-49.9$ & $-2.10$ & $-4.683$ & 1.0457 & 12.33 & 0.41 & 0.7 &    2.86 \\
  IC\_4651 & 5949561772753237248 & 260.920 & $-49.9$ & $-2.29$ & $-5.386$ & 1.2379 & 11.23 & 0.64 & 0.6 &   11.61 \\
  IC\_4651 & 5949565616715032064 & 261.128 & $-49.9$ & $-3.36$ & $-5.255$ & 1.0841 & 11.95 & 0.42 & 0.6 &   3.21 \\
  IC\_4651 & 5949553522087135616 & 261.231 & $-49.9$ & $-2.10$ & $-4.683$ & 1.0457 & 12.33 & 0.41 & 0.7 &    2.86 \\
  IC\_4651 & 5949553835654195968 & 261.273 & $-49.9$ & $-2.26$ & $-4.762$ & 1.1896 & 10.48 & 0.41 & 1.0 &   2.93 \\
%  IC\_4651 & 5949565616715032064 & 261.128 & $-49.9$ & $-3.36$ & $-5.255$ & 1.0841 & 11.95 & 0.42 & 0.6 &   3.21 \\
  IC\_4651 & 5949584170973723520 & 261.507 & $-49.7$ & $-2.02$ & $-5.612$ & 1.1522 & 12.32 & 0.56 & 0.6 &   17.13 \\
       \hline 
    \end{tabular}
\end{table*}

%-----------------------------------------------------------------------
\subsection{Table 3: Yellow stragglers in open clusters}
\label{subsec:third_table}
%-----------------------------------------------------------------------

This table lists information on every identified yellow straggler candidate. 
It contains eleven columns in the same format as in Table \ref{tab:example_BSS} (see \S~\ref{subsec:second_table} for more information).

%-----------------------------------------------------------------------
\subsection{Notes}
\label{subsec:notes}
%-----------------------------------------------------------------------

This file gathers information that clarifies, complements, or simply adds content to the Tables previously mentioned.

%-----------------------------------------------------------------------
\section{Statistics}
\label{sec:estadistica}
%-----------------------------------------------------------------------
\subsection{Comparison with AL95 and AL07}
\label{comparison}
%-----------------------------------------------------------------------

In total, 897 blue straggler candidates were identified in the 408 OCs investigated. The number of clusters with at least one blue straggler is 111 (27.20\%). 
In comparison with \citetalias{Ahumada_1995}, where 959 BS candidates were found in 390 clusters, and where the number of clusters with any stragglers was 225 (57.7\%), our percentage is $\sim 30$\% less. The respective numbers for \citetalias{Ahumada_2007} are as follows:  1887 stragglers, 427 clusters, and   199 clusters  with stragglers (46.6\%); here the difference is $\sim20$\%.
It can be safely  assumed that a good part of this  results from the greatly improved membership information available for the present work because of \emph{Gaia}. 
Regarding the YSS, 77 candidates were identified in 43 clusters (10.53\%),  with most of them hosting only one yellow straggler. These clusters and their  $N_\mathrm{YSS}$ are listed in Table~\ref{tab:clusters_ys}. 
No straggler candidates were found in clusters with  $\log(\mathrm{age}) \leq7.5$. In \citetalias{Ahumada_1995} and \citetalias{Ahumada_2007}, some stragglers -- albeit few -- were  identified  in young clusters; this  may reflect a bias towards bright magnitudes (see the case of   IC~2602 in \S~\ref{subsec:gaialimitaciones}), besides membership considerations.

\begin{table}
\centering
\caption{Open clusters harbouring YSS.}
\label{tab:clusters_ys}
\begin{tabular}{l c c | l c c} 
                \hline \hline
  Cluster  & $\log(\mathrm{age})$ & $N_\mathrm{YSS}$ & Cluster  & $\log(\mathrm{age})$ & $N_\mathrm{YSS}$\\
  \hline
 NGC~2477      &  8.80   &  4  &  Berkeley~69   &  8.95   &  1 \\
 Collinder~261 &  9.95   &  3  & King~7        &  8.82   &  1\\
 Berkeley~20   &  9.78   &  3  & Trumpler~20   &  9.11   &  1\\
 NGC~6253       &  9.70  &  3  & Berkeley~70   &  9.67   &  1\\
 NGC~2437      &  8.40   &  3  & NGC~5823      &  8.90   &  1\\
 Berkeley~19   &  9.40   &  3  & Berkeley~29   &  9.02   &  1 \\ 
 NGC~2141      &  9.23   &  3  & Pismis~3      &  9.02   &  1\\ 
 Ruprecht~75   &  9.15   &  3  & IC~1311       &  9.20   &  1\\
 King~5        &  9.10   &  3  & NGC~1245      &  9.00   &  1\\
 NGC~2158      &  9.02   &  3  & NGC~7142      &  9.55   &  1 \\
 Berkeley~23   &  8.90   &  3  & Melotte~66    &  9.53   &  1 \\
 Trumpler~5    &  9.60   &  2  & NGC~2243      &  9.03   &  1 \\ 
 Berkeley~39   &  9.90   &  2  & NGC~3680      &  9.07   &  1\\ 
 King~2        &  9.78   &  2  & NGC~2192      &  9.15   &  1\\
 Berkeley~32   &  9.70   &  2  & NGC~6705      &  8.40   &  1\\
 NGC~1193      &  9.70   &  2  & NGC~752       &  9.16   &  1\\ 
 NGC~2682      &  9.45   &  2  & IC~4651       &  9.05   &  1\\
 NGC~1798      &  9.25   &  2  & NGC~2354      &  9.18   &  1\\
 IC~166        &  9.09   &  2  & Berkeley~99   &  9.50   &  1\\
 NGC~2204      &  9.03   &  2  & NGC~3114      &  8.09   &  1 \\
 Tombaugh~2    &  9.01   &  2  & NGC~7654      &  8.20   &  1\\ 
 Berkeley~81   &  9.00   &  2  & \\
 \hline
\end{tabular}
\end{table}

\begin{table}
\centering
\caption{Open clusters with the largest absolute population of blue stragglers ($N_\mathrm{BSS}\geq10$) 
 in comparison with \citetalias{Ahumada_2007}.}
\label{tab:clusters_highBS}
\begin{tabular}{lccc} 
                \hline \hline
  Cluster  & $\log(\mathrm{age})$ & $N_\mathrm{BSS}$ & $N_\mathrm{BSS}(\mathrm{AL07})$\\
  \hline
 Trumpler~5    &  9.60   &  103   & 70  \\ 
 Collinder~261 &  9.95   &   53   & 54  \\
 NGC~6791      &  9.92   &   48   & 75  \\
 NGC~2158      &  9.02   &   39   & 40  \\
 Berkeley~18   &  9.63   &   32   & 126 \\
 Berkeley~32   &  9.70   &   27   & 37  \\
 NGC~1798      &  9.25   &   27   & 24  \\
 Tombaugh~2    &  9.25   &   27   & $\dots$  \\
 King~2        &  9.78   &   26   & 30  \\
 NGC~188       &  9.88   &   22   & 24  \\
 Berkeley~17   &  10.0   &   20   & 31  \\
 Berkeley~21   &  9.34   &   20   & 51  \\
 NGC~2141      &  9.23   &   18   & 24  \\
 King~11       &  9.04   &   18   & 27  \\
 Berkeley~39   &  9.90   &   18   & 43  \\
 NGC~7789      &  9.52   &   16   & 22  \\
 NGC~6819      &  9.36   &   15   & 29 \\
 NGC~6253      &  9.70   &   14   & 27  \\
 Melotte~66    &  9.53   &   14   & 35  \\
 NGC~2243      &  9.03   &   14   & 09  \\
 NGC~2506      &  9.00   &   14   & 15 \\
 Berkeley~12   &  9.60   &   13   & 15 \\
 Berkeley~19   &  9.40   &   13   & 01 \\
 NGC~1193      &  9.70   &   12   & 19 \\
 Berkeley~70   &  9.67   &   12   & 64  \\
 Trumpler~20   &  9.11   &   12   & $\dots$  \\ %preguntar
 NGC~2682      &  9.45   &   11   & 30  \\
 NGC~7142      &  9.55   &   11   & 37  \\
                \hline
        \end{tabular}
\end{table}

In Table~\ref{tab:clusters_highBS}, we list the clusters with $N_\mathrm{BS}\geq 10$. In apparent contrast to our remarks in the preceding paragraph, it is interesting to note that some of the OCs show a greater number of BSS candidates in our catalogue than  in AL07. For example, we found that Trumpler~5  has the largest straggler population  \citep{Rain_2020b}, but  AL07 listed only 70 stars, of which just four were classified as of `type~1' -- that is to say, bona~fide stragglers given the then available   membership information. 
The discrepancy may result from the different areas searched for BSS, a radius of $8'$ in AL07, and  $30'$ in the present work. 
However, for most of the clusters, the opposite occurs, and the number of catalogued BSS decreases considerably. Most of the objects in Table~\ref{tab:clusters_highBS}  were also targeted by AL07 as the richest in BSS. As for the age of these clusters, all of them are old ($\log(\mathrm{age}) \geq 9.0$), just as AL95 and AL07 found. 
A comparison between our catalogue and AL07 is possible only for those sources with RA and DEC information available. These were obtained from WEBDA\footnote{https://webda.physics.muni.cz/} or any data linked to ADS. For the sources we matched, we found large inconsistencies between our catalogue and AL07. First, for individual clusters, the percentage of BSS found to be non-members according to our criteria (\S~\ref{sec:miembros}) is about 10--60\% of the AL07 BSS. However, there are cases such as NGC~2477, whose AL07 BSS are all members, but that appear concentrated around the TO and sub-giant branch in the \emph{Gaia} CMD. On the other hand, for close and very well studied clusters with spectroscopically confirmed BSS, it is possible to retrieve about 70\% and in some cases up to 100\%, as \cite{Vaidya_2020} have demonstrated. We are aware that some BSS are lost given our stringent and conservative selection criteria. Although we did not attempt to estimate the number of missed BSS, we are confident that this is small because our choice of stars with $P_\mathrm{memb}\geq$ 50\% captures the majority of the cluster members, as found by other authors \citep{Cantat-Gaudin_2018, Carrera_2019, Banks_2019, Yontan_2019}.

%-----------------------------------------------------------------------
\subsection{$N_\mathrm{BSS}$/$N_\mathrm{MSS}$ versus\ age}
\label{subsec:BSMS_age}
%-----------------------------------------------------------------------

\begin{figure}
\centering 
\includegraphics[width=\linewidth]{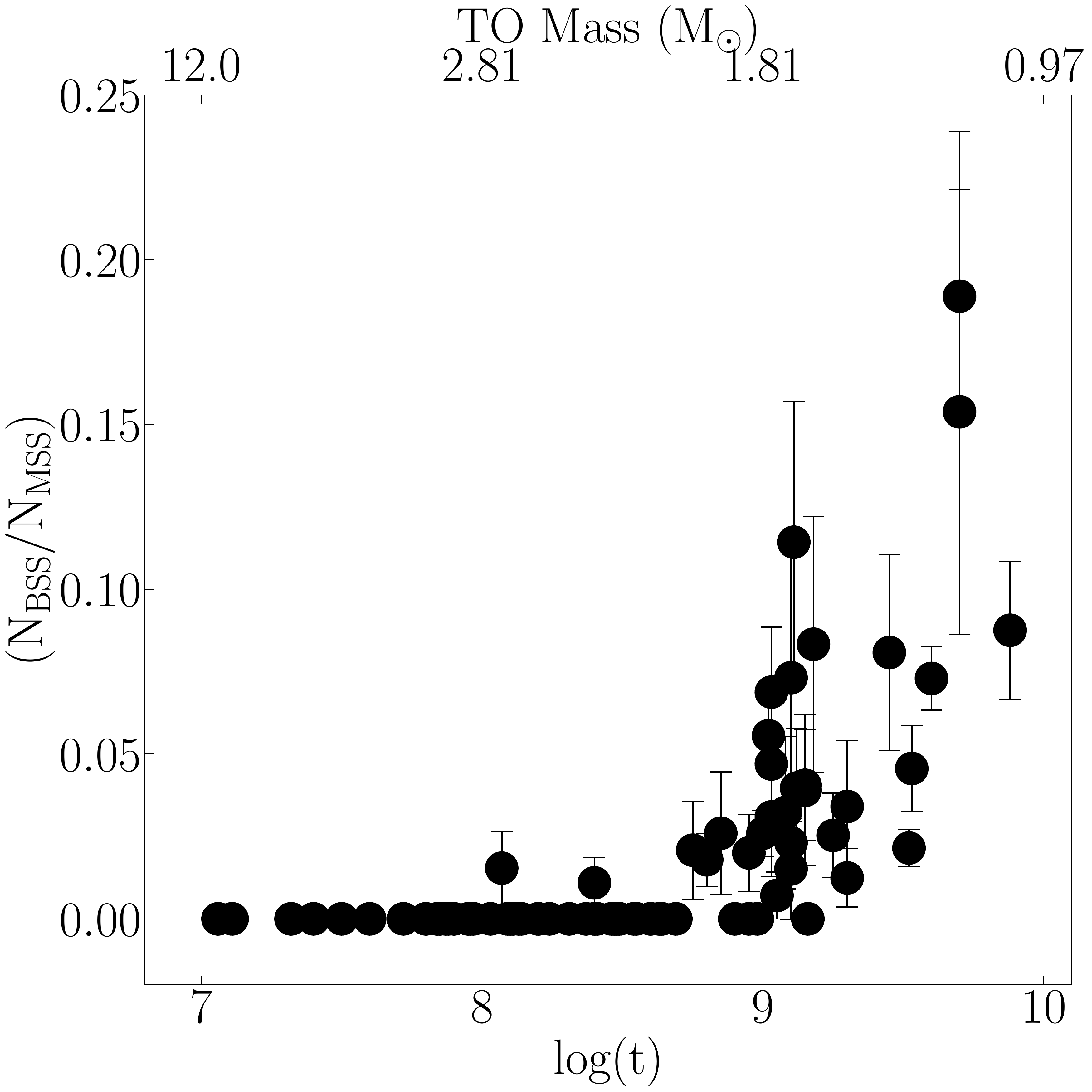}
\caption{N$_{BSS}$ /$N_\mathrm{MSS}$ as a function of the age in logarithmic scale. Only 107 clusters are included. Errors are Poisson.}
    \label{fig:relation_age_norm}
\end{figure}

Figure \ref{fig:relation_age_norm} shows the ratio $N_\mathrm{BSS}$/$N_\mathrm{MSS}$ versus cluster age, in a logarithmic scale, for 107 OCs. The $N_\mathrm{MSS}$ is the number of cluster main sequence stars up to 1 magnitude below the TO,  adopted as a proxy of the cluster richness. The clusters were chosen first by visually inspecting every CMD, built with stars of $P_\mathrm{memb} \geq 80\%$, and then by retaining those that showed well-defined  evolutionary phase signatures  -- that is main sequence, binary sequence, and red clump when it was possible -- and those that were well-matched to the corresponding isochrone.
In the top $x$-axis of the figure, the mass of the cluster TO for a given age is shown.  Star counts are  in Table~\ref{tab:clusters_ratio}. 
  
 \begin{table}
\caption{Clusters  with the largest relative population of blue stragglers ($N_\mathrm{BSS}/N_\mathrm{MSS} 
\geq 0.01$). Only stars with $P_\mathrm{memb} \geq 80\% $ were  considered.}
\label{tab:clusters_ratio}
\centering
\begin{tabular}{l c c c c} 
\hline \hline
  Cluster  & $\log(\mathrm{age})$ & $N_\mathrm{BSS}$ & $N_\mathrm{MSS}$  & $N_\mathrm{BSS}/N_\mathrm{MSS}$\\
  \hline
 Berkeley~32    &  9.70   &  17   & 90  & 0.19 \\
 NGC~6253       &  9.70   &  06   & 39  & 0.15 \\  
 Trumpler~20    &  9.11   &  08   & 70  & 0.11 \\
 NGC~188        &  9.88   &  19   & 217 & 0.09 \\ 
 NGC~2354       &  9.18   &  05   & 60  & 0.08 \\
 NGC~2682       &  9.45   &  08   & 99  & 0.08 \\
 King~5         &  9.10   &  03   & 41  & 0.07 \\
 Trumpler~5     &  9.60   &  62   & 850 & 0.07 \\
 NGC~2243       &  9.03   &  13   & 189 & 0.07 \\
 NGC~2158       &  9.02   &  28   & 504 & 0.06 \\
 NGC~2204       &  9.03   &  07   & 149 & 0.05 \\
 Melotte~66     &  9.53   &  13   & 285 & 0.05 \\
 Collinder~110  &  9.15   &  06   & 148 & 0.04 \\
 Melotte~71     &  9.12   &  05   & 126 & 0.04 \\
 NGC~2627       &  9.15   &  03   &  77 & 0.04 \\
 NGC~2420       &  9.30   &  03   &  88 & 0.03 \\
 NGC~6005       &  9.08   &  02   &  62 & 0.03 \\
 NGC~2660       &  9.03   &  03   &  97 & 0.03 \\
 NGC~2506       &  9.00   &  14   & 539 & 0.03 \\
 NGC~6940       &  8.85   &  02   &  77 & 0.03 \\
 NGC~2112       &  9.25   &  04   & 158 & 0.03 \\
 NGC~7044       &  9.10   &  06   & 260 & 0.02 \\
 NGC~7789       &  9.52   &  15   & 698 & 0.02 \\
 NGC~6134       &  8.95   &  03   & 150 & 0.02 \\ 
 NGC~2477       &  8.80   &  05   & 279 & 0.02 \\
 NGC~6067       &  8.07   &  02   & 130 & 0.02 \\
 NGC~3960       &  9.10   &  01   & 66  & 0.02 \\
 NGC~6603       &  8.30   &  01   &  96 & 0.02 \\
 NGC~6939       &  9.30   &  02   & 161 & 0.01 \\
 NGC~6705       &  8.07   &  02   & 183 & 0.01 \\
 IC~4651        &  9.04   &  01   & 141 & 0.01\\
\hline
        \end{tabular}\\
\end{table}

The results can be summarised as follows. The ratio is approximately constant for young OCs until $\log(\mathrm{age}) \sim 8.7$ ($\sim 500$~Myr), followed by a steep increase for older clusters. Previous works attempted to discern if this correlation between $N_\mathrm{BSS}/N_\mathrm{MSS}$ and the age is true and related with some intrinsic mechanism that would produce more stragglers in old clusters. Surprisingly, our results are more similar to AL95 than AL07 when the number of stragglers is normalised to main sequence stars. In AL07, the absolute number of stragglers grows with the cluster age starting from $\log(\mathrm{age}) \sim 6.5$, while in AL95, two trends are present, one is approximately constant for young clusters and the
second one grows with age.
One scenario to explain the difference observed between both figures is related to the lack of BSS in young OCs and the confusion of defining an accurate TO in the CMD, which means that it is hard to distinguish MS stars from BSS, misclassifying them as stragglers. Concerning the sudden increase,  it was proposed by AL07 that the number of stragglers observed in old OCs is a consequence of mechanisms such as mass transfer in close binaries. This remains an open issue that requires further investigation.

%-----------------------------------------------------------------
\section{Conclusions}
\label{sec:conclusions}
%-----------------------------------------------------------------
We have presented a new catalogue of blue and yellow stragglers in OCs, 
under the necessity  of updating the previous versions based on two important facts. First, the burst of all-sky and large surveys  over the whole range of the electromagnetic spectrum provides new,  homogeneous data for an enormous amount of objects,  allowing their study  over larger sky areas. 
Second, the fact that there is the need to identify  straggler candidates more reliably and accurately,  not merely based on their position in the CMD, but also using the  more accurate  astrometric membership provided by \emph{Gaia}. 
We thus defined homogeneous  criteria for the selection of the BSS  to derive proper-motion-cleaned BSS catalogues in all our OCs.
In this edition, 897 stragglers were identified in 111 clusters of a total of 408 OCs, of which 19 are recently discovered objects that are not present in the catalogue \citetalias{Ahumada_2007}; on the other hand, 39 clusters listed in AL07 are not present in this new version. Regarding yellow stragglers counts, 77 YSS were identified in 43 OCs. The  proper  motion decontamination allowed an unambiguous selection of YSS  in the CMD, which had not been   possible in previous  BSS studies in OCs, given the high field star contamination that the Galactic disc causes in the CMDs. We want to remark that this is the very first catalogue  containing this information.

We also want to draw attention to another important difference that the reader will find between this work and AL07: The classification of  straggler candidates in type 1 and type 2 defined by AL07 is  not included in our catalogue.   We instead give  the coordinates, the astrometric solution, and the distance to the cluster centre of each BSS and YSS. 

We hope that this new compilation will be a useful reference for, and give a new boost to, studies of blue straggler and yellow straggler populations in OCs. 
We also believe it can be of great use, particularly for new follow-up observations of radial velocities that will provide a strong confirmation of our selected straggler candidates' membership and improve our knowledge of these interesting objects.

%--------------------------------------------------------------
\begin{acknowledgements}
    MJ.~R. is supported by CONICY PFCHA through Programa de Becas de Doctorado en el extranjero- Becas Chile / \texttt{2018-72190617}
    G.C. acknowledges funding from Italian Ministry of Education,
    University and Research (MIUR) through the "Dipartimenti di
    eccellenza” project Science of the Universe.
\end{acknowledgements}

% WARNING
%-------------------------------------------------------------------
% Please note that we have included the references to the file aa.dem in
% order to compile it, but we ask you to:
%
% - use BibTeX with the regular commands:
%   \bibliographystyle{aa} % style aa.bst
%   \bibliography{Yourfile} % your references Yourfile.bib
%
% - join the .bib files when you upload your source files
%-------------------------------------------------------------------
\bibliography{bib}
\bibliographystyle{aa}
%\begin{thebibliography}{}

%  \bibitem[Baker(1966)]{baker} Baker, N. 1966,
%      in Stellar Evolution,
%      ed.\ R. F. Stein,\& A. G. W. Cameron
%      (Plenum, New York) 333

%   \bibitem[Balluch(1988)]{balluch} Balluch, M. 1988,
%      A\&A, 200, 58

%   \bibitem[Cox(1980)]{cox} Cox, J. P. 1980,
%      Theory of Stellar Pulsation
%      (Princeton University Press, Princeton) 165

%   \bibitem[Cox(1969)]{cox69} Cox, A. N.,\& Stewart, J. N. 1969,
%      Academia Nauk, Scientific Information 15, 1

%   \bibitem[Mizuno(1980)]{mizuno} Mizuno H. 1980,
%      Prog. Theor. Phys., 64, 544
   
%   \bibitem[Tscharnuter(1987)]{tscharnuter} Tscharnuter W. M. 1987,
%      A\&A, 188, 55
  
%   \bibitem[Terlevich(1992)]{terlevich} Terlevich, R. 1992, in ASP Conf. Ser. 31, 
%      Relationships between Active Galactic Nuclei and Starburst Galaxies, 
%      ed. A. V. Filippenko, 13

%   \bibitem[Yorke(1980a)]{yorke80a} Yorke, H. W. 1980a,
%      A\&A, 86, 286

%   \bibitem[Zheng(1997)]{zheng} Zheng, W., Davidsen, A. F., Tytler, D. \& Kriss, G. A.
%      1997, preprint
%\end{thebibliography}

\end{document}